\documentclass[%
 notitlepage,reprint,
 amsmath,amssymb,
 aps,prl
]{revtex4-1}

\usepackage{cases}
\usepackage{graphicx}
\usepackage{dcolumn}
\usepackage{bm}
\usepackage{color}
\usepackage{amsfonts,latexsym}
\usepackage{enumerate}
\usepackage{amsmath}
\usepackage{amssymb}
\usepackage{rotating}
\usepackage{float}

\newcommand{\mi}{\mathrm{i}}

\newcommand{\dif}{\mathrm{d}}

\newcommand{\Eq}[1]{Eq.~\eqref{#1}}

\newcommand{\Fig}[1]{Fig.~\ref{#1}}

\newcommand{\beq}{\begin{equation}}
\newcommand{\eeq}{\end{equation}}

\newcommand{\beqa}{\begin{eqnarray}}
\newcommand{\eeqa}{\end{eqnarray}}

\def\bal#1\eal{\begin{align}#1\end{align}}

\def\Bal#1\Eal{\begin{align*}#1\end{align*}}

\newcommand{\Beqa}{\begin{eqnarray*}}
\newcommand{\Eeqa}{\end{eqnarray*}}

\newcommand{\nn}{\nonumber}

\newcommand{\vect}[1]{\mathbf{#1}}

\begin{document}
\title{Three-body scattering hypervolumes of particles with short-range interactions}
\author{Shangguo Zhu$^{1}$}
   \email{shangguo.zhu@gatech.edu}
\author{Shina Tan$^{1,2}$}
    \email{shina.tan@physics.gatech.edu}
    \affiliation{$^{1}$School of Physics, Georgia Institute of Technology, Atlanta, Georgia 30332, USA}
    \affiliation{$^{2}$Center for Cold Atom Physics, Chinese Academy of Sciences, Wuhan 430071, China}

\begin{abstract} 
The low-energy scattering of three bosons or distinguishable particles with short-range interactions is characterized by a fundamental parameter,
the three-body scattering hypervolume. Its imaginary part is directly related to the three-body recombination rate in a quantum gas consisting of such particles.
We derive an analytical formula of it for weak interactions,
and perform its first numerical calculations for bosons with a variable nonzero-range potential.
For attractive interactions, we identify several three-body resonances at which the three-body scattering hypervolume becomes divergent or anomalously large. 
\end{abstract}
\maketitle


\textbf{Introduction.} The three-body scattering hypervolume $D$ is a fundamental parameter characterizing the effective three-body interaction~\cite{Tan2008}, in analogy to the two-body scattering length $a$ characterizing the two-body interaction~\cite{Dalibard1999}. It is important in the physics of three or more bosons or distinguishable particles, such as ultracold atomic gases~\cite{Tan2008}, multi-meson systems~\cite{Beane2007,Beane2011}, halo nuclei consisting of three loosely bound subsystems~\cite{Frederico2012}, \textit{etc}. At $a=0$, the ground state energy per particle of a dilute Bose-Einstein condensate (BEC) is 
\beq 
\label{BEC} 
E/N= \hbar^2  n^2 D/6m + o(n^2), 
\eeq 
where $\hbar$, $m$, $N$ and $n$ are the Planck constant over $2\pi$, the mass of one boson, the number of bosons, and the number density, respectively~\cite{Tan2008}. 


The three-body scattering hypervolume $D$ has been revealed for only a couple of bosonic models. 
For bosons with large scattering lengths, 
Braaten, Hammer, and Mehen calculated the three to three scattering coupling constant $g_3^{}$ in the effective field theory, from which $D$ can be inferred~\cite{Braaten2002, Tan2008}. 
In 2008, $D$ of hard-sphere bosons was found to be $D_{\rm HS}\approx
1761.5 a^4$~\cite{Tan2008}. For most systems, the knowledge of this fundamental parameter is lacking.

When the interaction supports two-body bound states, after three particles collide, two of them may form a dimer (a two-body bound state). The dimer and the third particle kinetically gain the released binding energy. This process is known as the three-body recombination, which causes atom loss in dilute ultracold gases~\cite{Soding1999,Stenger1999,Roberts2000}. The measurement of the three-body recombination rate constant $L_3^{}$ serves as a probe of few-body phenomena, such as the Efimov effect~\cite{Efimov1970a,Efimov1970,Kraemer2006a}. $L_3^{}$ for atoms with large $|a|$ has been calculated~\cite{Esry1999,Bedaque2000,Braaten2001,Braaten2006}. Meanwhile, the three-body recombination contributes a negative imaginary part to the energy, and this suggests a complex $D$ according to \Eq{BEC}.


In this letter, we investigate the three-body scattering hypervolume analytically and numerically. First, in the limit of weak two-body interactions,
we find that $D$ depends on the interaction strength quadratically. Second, when the interaction supports two-body bound states we find that $D$ becomes complex. Its imaginary part is directly related to $L_3^{}$ and is expressed as a sum of the contributions from different dimer states. This provides a novel method to calculate $L_3^{}$.
Third, we perform the first numerical calculations of $D$ for bosons with a variable nonzero-range interaction. Our results are consistent with the previous result for hard-sphere bosons in
the strong-repulsion limit and the analytical formula we derived in the weak interaction limit. For attractive interactions, by sweeping the depth of the potential,
we identify three-body resonances.


The three-body scattering hypervolume $D$ is defined in either of the two asymptotic expansions of the wave function of three identical bosons colliding at zero energy, zero total momentum and zero total orbital angular momentum~\cite{Tan2008}. One expansion, called the 111-expansion, is when the three pair-wise distances go to infinity simultaneously; the other, called the 21-expansion, is when the distance between two particles stays fixed and the distance between the third particle and the pair goes to infinity~\cite{expTan2008}. The leading order of the 111-expansion is assumed to be $1$, which corresponds to the most important incoming three-body channel and fixes the wave function uniquely. $D$ appears as a parameter in the $1/\rho^4$ or $1/y_i^4$ term in the 111- or 21- expansion, respectively. Here we have defined the Jacobi vectors $\vect{x}_i^{} =\vect{r}_j^{}-\vect{r}_k^{}$ and $\vect{y}_i^{} =\frac{2}{\sqrt{3}}[\vect{r}_i^{}-\frac{1}{2}(\vect{r}_{j}^{}+\vect{r}_k^{})]$, where $(ijk)$ is an even permutation of (123) and $\vect{r}_1^{}$, $\vect{r}_2^{}$, $\vect{r}_3^{}$ represent the position vectors of the three bosons, and $\rho$ is the hyperradius, defined as $\rho\equiv\sqrt{(x_1^2+x_2^2+x_3^2)/2}=\frac{\sqrt3}{2}\sqrt{x_i^2+y_i^2}$.


\textbf{Weak interactions.} We analytically derive an approximate formula of $D$ when the interaction is weak. One can express the wave function in Born series $\psi = \psi_0^{}+ G\mathcal{V} \psi_0^{} +(G\mathcal{V})^2 \psi_0^{} +\cdots$, where $\psi_0^{}=1$ is the solution to the free Schr\"odinger equation, $G$ is the Green's operator ($G=-T^{-1}$, where $T$ is the kinetic energy operator), and $\mathcal{V}$ is the potential energy operator. If the interaction contains two-body potentials only, namely $\mathcal{V}= V(x_1^{})+V(x_2^{})+V(x_3^{})$, where $V(x)$ is a short-range potential whose characteristic range is $r_0^{}$, we find that when $x_1, x_2, x_3 \gg r_0^{}$, so that the two-body potentials vanish,
$G\mathcal{V} \psi_0^{} = \sum_{i=1}^{3} -\alpha_1^{}/x_i^{}$, and 
\beq
(G\mathcal{V})^2 \psi_0^{} = \big(\sum\limits_{i=1}^{3}\frac{\beta}{ x^{}_i}+\frac{4\sqrt{3}\theta_i^{} \alpha_1^{2}}{\pi \rho^2\sin 2\theta_i^{}} \big)
-\frac{\sqrt{3} \alpha_1^{} \alpha_3^{}}{\pi \rho^4}+O\big(\frac{1}{\rho^6}\big), \nn
\eeq
where  $\theta_i^{} = \arctan(y_i^{}/x_i^{})$, $\alpha_n^{}= \frac{m}{\hbar^2}\int_{0}^{\infty}\dif x~ x^{n+1}V(x)$, and $\beta = \frac{m^2}{\hbar^4} \int_{0}^{\infty}\dif x \int_{0}^{x}\dif x'~ 2 x x'^2 V(x) V(x')$. By comparison with the 111-expansion, we find the leading order expression of $D$:
\beq
\label{weak}
D=8\pi^2\alpha_1^{} \alpha_3^{}+O(V^3),
\eeq
and $a=\alpha_1-\beta+O(V^3)$.
We see that $D$ depends on the strength of the interaction quadratically. The next-to-leading order correction is proportional to the cube of the strength of interaction. 


\textbf{Complex $D$ and three-body recombination.} The original 21-expansion is applicable when the interaction does not support any two-body bound state~\cite{Tan2008}. In this case, no three-body recombination occurs and $D$ is real. When the interaction is attractive and strong enough to support two-body bound states, three bosons may recombine into a dimer and a free boson, which carry the released binding energy and behave as an outgoing wave. Then, the three-body wave function becomes
\beq
\label{wavefunction}
\phi = \phi_0^{} + \sum_{l=0}^{l_\text{max}}\sum_{\nu=0}^{\nu_l^{}}C_{l\nu}^{} \sum_{i=1}^{3}\phi_{l\nu}^{}(\vect{x}_i^{}, \vect{y}_i^{}),
\eeq
where $\phi_0^{}$ is the incoming wave plus the elastically scattered wave.
$\phi_0^{}$ obeys the original 111- and 21- expansions~\cite{Tan2008}. $\phi_{l\nu}^{}(\vect{x}_i^{}, \vect{y}_i^{})$ is the outgoing wave in the $l\nu$-channel, where $l$ and $\nu$ are the orbital angular momentum quantum number and the vibrational quantum number of the dimer, respectively. $\nu_l^{}$ is the maximum $\nu$ at a particular $l$,
and $l_\text{max}$ is the maximum $l$ for the dimer states. At large  $y$, such that the free boson is outside the range of interaction
with the other two bosons, $\phi_{l\nu}^{}$ can be expressed as 
\beq
\label{bswf}
\phi_{l\nu}^{}(\vect{x}, \vect{y})= u_{l\nu}^{}(x) \mi^{l+1} h_l^{(1)}(\kappa_{l\nu}^{} y) P_l^{}(\hat{\vect{x}}\cdot \hat{\vect{y}}),
\eeq
where $P_l^{}$ is the Legendre polynomial, $h_l^{(1)}$ is the spherical Hankel function of the first kind, and $\kappa_{l\nu}^{}>0$ is the binding wave number defined such that the binding energy of the dimer is $\hbar^2\kappa_{l\nu}^{2}/m$. $u_{l\nu}^{}(x)$ is the radial part of the dimer wave function, satisfying the two-body Shr\"odinger equation and the normalization condition
\bal
[-\frac{\hbar^2}{m}\nabla_x^2+V(x) +\frac{\hbar^2}{m} \kappa_{l\nu}^{2}]u_{l\nu}^{}(x)P_l^{}(\hat{\vect{x}}\cdot \hat{\vect{y}})= 0, \\
\int_{0}^{\infty}\dif x~ x^2 u_{l\nu}^{\ast}(x) u_{l\nu'}^{}(x) = \delta_{\nu\nu'}^{}(2l+1)/4\pi \kappa_{l\nu}^{3} .
\eal

As the outgoing wave contributes to a positive probability flux towards the outside of a large hyperspherical surface, $D$ gains a negative imaginary part to make the net flux vanish and conserve the probability. Then, a relation between the imaginary part of $D$ and the coefficients $C_{l\nu}^{}$ is established
\beq
\label{ImD}
{\rm Im} (D) = -\frac{9\pi\sqrt{3}}{2} \sum_{l\nu}\frac{|C_{l\nu}^{}|^2}{\kappa_{l\nu}^4},
\eeq
where the summation is over all the dimer states, and the symbol ${\rm Im}$ denotes the imaginary part.

When $D$ becomes complex, the ground state energy in \Eq{BEC} gains a negative imaginary part as well, indicating the decaying of the BEC. At a short time $t$, the probability
that no recombination occurs is $\exp(- 2 |{\rm Im}(E)| t/ \hbar)\approx1-2 |{\rm Im}(E)| t/ \hbar$. Then, the probability for one recombination is $2 |{\rm Im}(E)| t/ \hbar$. After each recombination, three atoms escape from the trap. This leads to the formula of the three-body recombination rate constant
\beq
\label{3bloss}
L_3^{}= -\frac{\hbar}{ m} {\rm Im}(D),
\eeq
where $L_3^{}$ is defined as $\dif n/\dif t= -L_3^{} n_{}^3$. According to the relation between $D$ and the three to three scattering coupling constant $g_3^{}$, this is consistent with the effective field theory formulation~\cite{Tan2008,Braaten2002}. Together with \Eq{ImD}, the number of the produced dimers in each channel can be calculated given the coefficient $C_{l\nu}^{}$
and the binding wave number $\kappa_{l\nu}^{}$. The distribution of the numbers of different dimer products has been measured  experimentally for ultracold Rb atoms~\cite{Wolf2017}. It was found that the propensity rule of the contribution is roughly the inverse square root of the dimer binding energy. However, from the following numerical calculations for bosons with Gaussian interaction potentials, we find that the contribution does not have a monotonic dependence on the binding energy.



\begin{figure*}[ht]
	\includegraphics[scale=0.7]{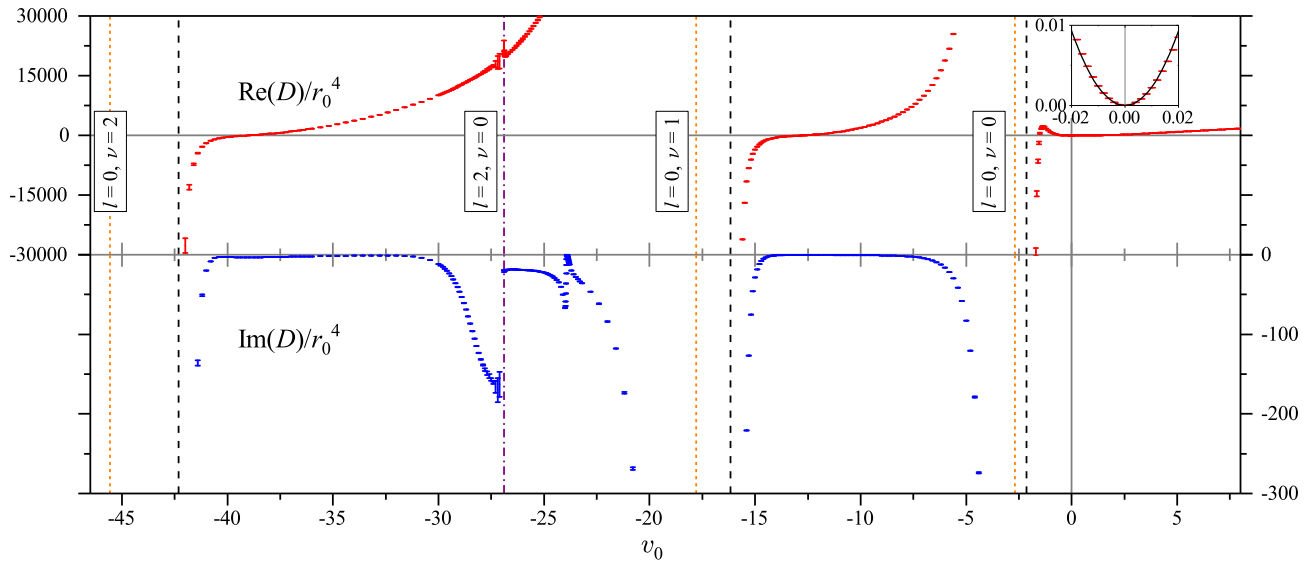}
	\caption{\label{fig:overall} The three-body scattering hypervolume $D$ in units of $r_0^4$ as a function of the dimensionless strength $v_0^{}$ of the potential. The red (blue) error bars in the upper (lower) chart represent the real (imaginary) part of $D$. The vertical orange dotted (purple dotdashed) lines at $v_0\approx -2.6840, -17.796$, and $-45.574$ ($v_0^{} \approx -26.901$) indicate where the \textit{s}-wave (\textit{d}-wave) bound states emerge. The vertical black dashed lines at $v_0\approx -2.1308, -16.163$, and $-42.32$ indicate where the three-body resonances occur. The inset shows $D$ at small $v_0$, where the solid line represents the leading order result in \Eq{weakG}. } 
\end{figure*}

\textbf{Numerical calculation.} For a general pair-wise short-range interaction, we need to solve the Shr\"odinger equation numerically to extract $D$ and the coefficients $C_{l\nu}^{}$. At a large hyperadious $\rho$, the wave function $\phi$ can be approximated by the outgoing wave function and the asymptotic 111- and 21-expansions according to \Eq{wavefunction}. We can treat it as the Dirichlet boundary condition on a large hyperspherical surface with $\rho=\rho_c^{}$. Then, the wave function inside the hypersphere is uniquely determined. The unknown $D$ and $C_{l\nu}^{}$ can be fixed by making the hyperradial derivative of the wave function continuous along the hyperradial direction. 

Note that the 111- and 21-expansions do not have the same error $O(1/\rho_c^8)$ uniformly on the whole hyperspherical surface~\cite{Tan2008}. When $x_i^{} \sim \rho_c^{1/3}$ and $y_i^{} \sim \rho_c$  both expansions have errors $O(1/\rho_c^6)$. We may reduce the error by obtaining more high order terms in each expansion. However, another three-body parameter, different from $D$, will appear in the $1/\rho^8$ or $1/y_i^8$ term. Without introducing any other three-body parameter, we improve the 111-expansion by adding only a portion of the higher order terms, such that all the terms in the 21-expansion are included by the improved 111-expansion in their common region $y_i^{}\gg x_i^{} \gg r_0^{}$. The improved 111-expansion then has error $O(1/\rho_c^8)$ whenever the three bosons are outside the range of interaction,
and we can calculate $\phi_0$ with error $O(1/\rho_c^8)$ everywhere on the hyperspherical surface.

In the following, we consider a model potential
\beq
V(\vect{r})=v_0^{}\frac{\hbar^2}{m r_0^2} \exp\Big(-\frac{r^2}{r_0^2}\Big),
\eeq
where the dimensionless parameter $v_0^{}$ is variable. When $v_0^{}$ is large and positive, the interaction approaches the hard-sphere limit. When $v_0^{}$ is negative and large enough, it supports two-body bound states. At $v_0\approx -2.6840, -17.796$, and $-45.574$, \textit{s}-wave bound states emerge, labeled as $(l=0, \nu=0)$, $(l=0, \nu=1)$ and $(l=0, \nu=2)$, respectively. At these critical values, $a$ diverges, and there are an infinite number of Efimov states~\cite{Efimov1970,Efimov1970a}. At $v_0\approx -26.901$, a \textit{d}-wave bound state emerges, labeled as $(l=2, \nu=0)$, and the \textit{d}-wave scattering ``length" $a_d^{}$ (with dimension $[{\rm length}]^5$) diverges. We say that the system is at the two-body \textit{s}-wave or \textit{d}-wave resonance if $a$ or $a_d$ diverges.
Experimentally this can be realized by Feshbach resonances~\cite{Chin2010}.

We calculate $D$ numerically for various $v_0^{}$ and plot the results in \Fig{fig:overall}. Our results have errors due to a finite number of basis functions that we used to expand the wave function on the hyperspherical surface, and a finite $\rho_c$. The errors become larger when the size of the dimer or the two-body parameters like $a$ and $a_d^{}$ increase. So the calculation fails when we are too close to the two-body resonances. The maximum value of $|a|$ is limited to be of the same order of $r_0^{}$, and therefore we do not have numerical results in the Efimov regimes in which $|a|\gg r_0^{}$.

\begin{figure*}[ht]
	\includegraphics[scale=0.7]{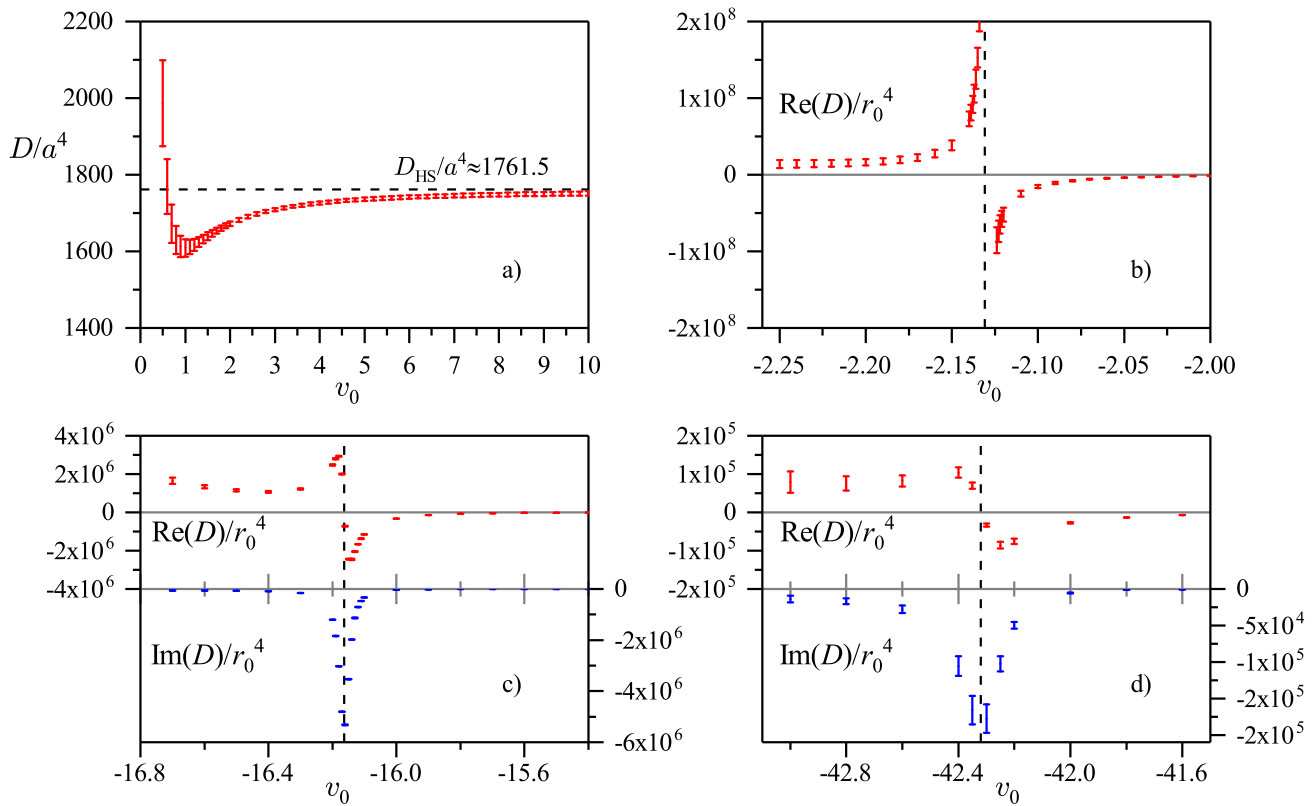}
	\caption{\label{fig:res} a) displays $D/a^4$ as a function of $v_0^{}$ when $v_0^{}>0$. The horizontal dashed line represents the value for hard-sphere bosons. b), c), and d) show	$D$ in units of $r_0^4$ as a function of $v_0^{}$ near three-body resonances at $v_0\approx -2.1308$, $-16.163$, and $-42.32$, respectively. In a) and b), ${\rm Im}(D)$ vanishes.}
\end{figure*}

When $|v_0^{}|$ is small, we can use \Eq{weak} to obtain
\beq
\label{weakG}
D/r_0^4 =\frac{3}{4}\pi^3 v_0^{2}+ O(v_0^3).
\eeq 
From the inset of \Fig{fig:overall}, we see that the numerical results agree with this analytical formula. From \Fig{fig:res}a, we see that $D$ approaches the hard-sphere limit $D_{\rm HS}/a^4\approx
1761.5$ when $v_0^{}$ is positive and large~\cite{Tan2008}.



We find several three-body resonances at $v_0<0$. At $v_0^{}\approx-2.1308$, $D$ has a simple pole shown in \Fig{fig:res}b. As there is no two-body bound state, the imaginary part of $D$ is zero here. We find an approximate formula near the pole: $D/r_0^4\approx (\frac{-6.2}{v_0^{}+2.1308} +56) \times 10^5$. As $D$ characterizes the effective three-body interaction, the pole indicates a three-body resonance. When $v_0^{}$ is slightly less than $-2.1308$, $D$ becomes large and positive and a shallow three-body bound state emerges, as can be verified by a numerical solution of the Schr\"{o}dinger equation at nonzero energy. Such a shallow three-body bound state is a Borromean state \cite{Frederico2012} because any two of the particles are not bound, and it is qualitatively similar to some nuclear halo states \cite{Frederico2012}.
At the pole, $a\approx-4.38 r_0^{}$, which is not much larger than $r_0$, so this three-body bound state is different from an Efimov state. It is stable because there is no dimer state that it can decay into. In the region $v_s^{(0)}<v_0^{}<-2.1308$,  where $v_s^{(0)}\approx-2.6840$,
$D$ should have an infinite number of simple poles with the accumulation point at $v=v_s^{(0)}$ due to the Efimov effect.

We show two more three-body resonances at $v_0^{}\approx -16.163$ and $v_0^{}\approx -42.32$ in \Fig{fig:res}c and \Fig{fig:res}d, respectively. We see that ${\rm Re}(D)$ experiences a sharp transition from a large negative peak to a large positive peak as we decrease $v_0^{}$. Meanwhile, ${\rm Im}(D)$ has a sharp dip right at the resonance, indicating rapid three-body recombination. Near each of these resonances we expect a shallow metastable three-body state, which decays due to three-body recombination or three-body dissociation.

We find some other prominent features in \Fig{fig:overall}. Near the \textit{d}-wave resonance at $v_0=v_d^{(0)}\approx -26.901$, we find an upward peak for ${\rm Re}(D)$ 
as we approach $v_d^{(0)}$ from above. We do not know whether $\text{Re}(D)$ diverges or not at $v_0=v_d^{(0)}$. We also find that ${\rm Im}(D)$ exhibits a possible discontinuity at $v_0=v_d^{(0)}$. 

At $v_0^{}\approx -23.9$ where $a\approx 2.24 r_0^{}$, $|{\rm Im}(D)|$ reaches a local minimum $\approx 0.194(10)$, indicating a local minimum of the three-body recombination rate. A similar pattern is seen when $a$ is large and positive, due to the destructive interference of competing pathways~\cite{Esry1999}.

\textbf{Summary.} 
We have investigated the three-body scattering hypervolume $D$ analytically and numerically. 
For weak two-body interactions, we have derived an approximate formula for $D$. 
We extended the concept of $D$ to the complex plane when the interaction supports two-body bound states. We developed a numerical method for calculating $D$ for nonzero-range interactions. We numerically calculated $D$ for bosons with pair-wise Gaussian potentials; when the potential is strongly repulsive or weak, it agrees well with the hard-sphere result or the weak interaction formula, respectively; when the potential is sufficiently attractive, we have identified several three-body resonances.


\begin{acknowledgments}
	We thank Chris Greene, Paul Julienne, Doerte Blume, Kevin Driscoll, and Kevin Daily for discussions. We gratefully acknowledge support by the National Science Foundation CAREER Award Grant No. PHY-1352208. This research was supported in part by the National Science Foundation under Grant No. NSF PHY-1125915. We also thank the hospitality and support of the Kavli Institute for Theoretical Physics at Santa Barbara during the program ``Universality in Few-Body Systems". 
\end{acknowledgments}


\end{document}